\documentclass[12pt,preprint]{aastex}

\slugcomment{Submitted to the Astronomical Journal}

\shorttitle{FUSE Spectroscopy of SS Aur and RU Peg}
\shortauthors{Sion et al.}

\begin{document}

\title{Far Ultraviolet Observations of the Dwarf Novae SS Aur and RU Peg
       in Quiescence} 

\author{Edward M. Sion, Fuhua Cheng, Patrick Godon\altaffilmark{1}, 
Joel A. Urban}
\affil{Astronomy and Astrophysics, Villanova University, \\ 
800 Lancaster Avenue, Villanova, PA 19085, USA}
\email{patrick.godon@villanova.edu edward.sion@villanova.edu, 
       fcheng@ast.vill.edu, jurban@ast.vill.edu}

\author{Paula Szkody}
\affil{Department of Astronomy,
University of Washington,
Seattle, WA 98195, USA}
\email{szkody@alicar.astro.washington.edu}

\altaffiltext{1}{Visiting at the 
Space Telescope Science Institute, Baltimore, MD 21218, USA}

\clearpage 

\begin{abstract}

We have analyzed the Far Ultraviolet Spectrocopic Explorer (FUSE)  
spectra of two U Gem-Type dwarf novae, SS Aur and RU Peg, 
observed 28 days and 60 days (respectively) after their last outburst. 
In both systems the FUSE
spectra ($905 - 1182$ \AA)  reveal evidence of the underlying accreting
white dwarf exposed in the far UV. Our grid of theoretical models yielded
a best-fitting photosphere to the FUSE spectra with
T$_{eff}$=31,000K for SS Aur and  T$_{eff}$=49,000K for RU Peg.  
This work provides two more dwarf nova systems with known 
white dwarf temperatures above the period gap where few are known. 
The absence of C\,{\sc iii} (1175 \AA) absorption in SS Aur and 
the elevation of N above solar suggests the possibility that SS
Aur represents an additional accreting white dwarf where the surface C/N
ratio derives from CNO processing. For RU Peg, the modeling uncertainties
prevent any reliable conclusions about the surface abundances and
rotational velocity.

\end{abstract}


\newpage 

\section{Introduction: Accretion Onto White Dwarfs in Dwarf Novae} 

Dwarf Novae (DN) are a subclass of Cataclysmic Variable (CV) Systems, in
which a white dwarf (WD, the primary) accretes hydrogen-rich matter from a
low-mass main sequence-like star (the secondary) filling its Roche lobe.
The transferred gas forms an accretion disk around the white dwarf,
and is subject to a
thermal instability that causes cyclic changes of the accretion rate. A
low rate accretion ($\approx 10^{-11} M_{\odot} yr^{-1}$) quiescence stage
is followed every few weeks to months by a high rate accretion ($\approx
10^{-8} M_{\odot} yr^{-1}$) outburst stage of days to weeks. It is believed
that these
outbursts (dwarf nova accretion event or nova-like high state), are
punctuated every few thousand years or more by a Thermonuclear Runaway
(TNR) explosion: the classical nova \citep{war95}. \\

In some dwarf novae during quiescence, the
accretion rate and disk temperature has so greatly declined that the disk 
emits mainly in the optical and has little or no contribution 
to the FUV spectrum. Consequently, the
underlying hot white dwarf accreter is exposed spectroscopically and
dominates the far UV light. We have been studying these systems
extensively with HST since we are able to directly observe the physical
effects of disk accretion onto the white dwarf.  These natural
\textit{accretion
laboratories} have yielded, for the central accreters, the first
rotation
rates (due to accretion spinup), $T_{eff}$'s, cooling rates, chemical
abundances of the accreted plus original white dwarf matter, and dynamical
and gravitational redshift masses. Unfortunately, the number of such
systems above the period gap has been very limited (U Gem and
RX And are the only solid cases thus far). It is of considerable interest
to find other such systems.

Above the gap, dwarf novae even during quiescence may have accretion rates
sufficiently high that an optically thick disk and/or very hot boundary
layer dominate their FUV spectra. Moreover, these higher
accretion rates would also be expected to produce significant long term
heating of the white dwarf accretor. However, 
for the disk instability model to
apply as the mechanism for their dwarf nova outbursts, the mean accretion rate
$\dot{M}$ must be lower than the critical rate $M_{crit}$ 
to trigger the instability. 
Systems with mean $\dot{M} > M_{crit}$ do not exhibit cyclic changes 
of $\dot{M}$ \citep{sha86}.  
An obvious fundamental observational question is:
which hot component dominates in the far UV and what are its physical
properties?

Among the high $\dot{M}$ U Gem-type systems, two objects, SS Aur and
RU
Peg, have extensive earlier multi-wavelength studies and reasonably
well-constrained distances to represent important test cases. Their dwarf
nova behavior parameters are listed in Table 1.

RU Peg has an orbital period $P_{orb}$ = 0.3746 days \citep{sto81}, a system
inclination $i = 33^{o}$ \citep{sha83}, a secondary (mass donor)
spectral type K2-5V \citep{wad82,sha83}, a primary (WD) mass
$M_{wd} = 1.29\pm0.20 M_{\odot}$ \citep{sha83}. The near-Chandrasekhar
mass for the white dwarf has been corroborated by the Sodium (8190\AA)
doublet radial velocity study of \citet{fri90}.  They obtained a mass of
1.24 M$_{\odot}$ for the white dwarf and also found very good agreement
with the solution, including agreement with the range of plausible
inclination, found in the study by \citet{sto81}. Recently, a Hubble FGS
parallax of $3.55\pm0.26$ mas was measured by \citet{joh03}.

SS Aur has an orbital period $P_{orb}$ = 0.1828 days 
\citep{shahar86}.  The mass of
its white dwarf has been estimated 
to be $M_{wd} = 1.08\pm0.40$ $M_{\odot}$, while
its secondary is known to have a mass 
$M_{2} = 0.39\pm 0.02$ $M_{\odot}$, and
a system inclination $i = 38^{o}\pm16^{o}$ \citep{sha83}. SS Aur
has an
HST FGS parallax measurement of 497 mas \citep{har99}.

In two recent studies, \citet{lak01} and \citet{sio02} analyzed IUE far UV
archival spectra of both systems which revealed they contain very hot
white dwarfs. For SS Aur, their best-fit model photosphere has T$_{eff} =
30,000$K, log $g$ = 8.0, and solar composition abundances, while the
best-fit
accretion disk model has M$_{wd} = 1.0$ $M_{\odot}$, $i = 41^{o}$, and
$\dot{M} = 10^{-10}$ $M_{\odot}yr^{-1}$. They used a measured parallax
(497mas; 201 pc) for SS Aur, from observations with the HST FGS \citep{har99},
together with the scale factor S $= 1.12\times 10^{-3}$ from their best
fit, to calculate a radius for the emitting source of $4.68\times 10^{8}$ 
cm. This radius is clearly that of a compact object, a white dwarf, not
an accretion disk.

For RU Peg during quiescence, their best-fit high gravity solar
composition photosphere models yielded T$_{eff} = 50 - 53,000$K with scale
factor distances of $\sim$ 250 parsecs. Optically thick accretion disk
models
imply accretion rates between $1\times 10^{-9} M_{\odot}yr^{-1}$ and
$1\times
10^{-10} M_{\odot}yr^{-1}$ in order to match the steeply sloping far
UV
continuum, but yielded distances of 600 to 1300 parsecs, well beyond the
estimated distance range of 130 to 300 parsecs (now known to be 282 pc
from the new parallax by \citet{joh03}). However, they could
not
rule out that the far UV energy distribution is due to a multi-temperature
white dwarf with cooler, more slowly rotating higher latitudes and a
rapidly spinning, hotter equatorial belt.

Is the FUV spectral energy distribution best represented by a model of an
accretion disk alone, a composite white accretion disk plus white dwarf, a
rapidly spinning accretion belt and high gravity photosphere with an
inhomogeneous temperature distribution or by a uniform temperature white
dwarf synthetic spectrum alone? We will make use of the parallaxes
for both objects to determine the source of the FUV energy distribution.
FUSE offers a large variety and broad range of critical line transitions
at high spectral resolution across a broad range of ionization
states/levels and elements to help us advance our understanding of the
physics of the boundary layer and accretion.  

\section{Observations and Analysis} 

FUSE is a low-earth orbit satellite, launched in June 1999. Its optical
system consists of four optical telescopes (mirrors), each separately
connected to a different Rowland spectrograph. The four diffraction
gratings
of the four Rowland spectrographs produce four independent spectra on
two photon
counting area detectors. Two mirrors and two gratings are coated
with SiC to provide wavelength coverage below 1020 \AA, while the other
two mirrors and gratings are coated with Al and a LiF overcoat.  The
Al+LiF coating provides about twice the reflectivity of SiC at
wavelengths $>$1050 \AA, and very little reflectivity below 1020 \AA\
(hereafter the SiC1, SiC2, LiF1 and LiF2 channels). \\

A spectrum of SS Aur in quiescence was taken by FUSE on February 13, 2002
at 07:01 UT (MJD52318) approximately 28 days after the last outburst. The
exposure time was 14,513 seconds through the low resolution (LWRS:
30"x30") aperture. A spectrum of RU Peg in quiescence was taken by FUSE on
July 4, 2002, at 17:09 UT (MJD52459) approximately 60 days after the
last outburst. The exposure time was 1060 seconds through the LWRS
aperture. LWRS was used in both cases since it is
least prone to slit losses due to the misalignment of the four FUSE
telescopes.
The calculated S/N of the co-added spectra at $0.1$ \AA\
resolution 
is 3.65 for RU Peg and 4.67 for SS Aur. The S/N improves to $\approx 10$
at $0.5$ \AA\ resolution. 
It is clear that the relatively poor FUSE spectral quality of
both SS Aur and RU Peg speaks to the requirement for more observing time.
For example, SS Aur should have had at least 20,000 sec while for RU
Peg, at least 5,000 sec is required.  

The data used were reduced with the CalFUSE pipeline version 2.1.6.
In this version, event bursts are automatically taken care off. 
Event bursts are short periods during an exposure when high count 
rates are registered on one of more detectors. The bursts exhibit  
a complex pattern on the detector, their cause, however, is yet unknown 
(it has been confirmed that they are not detector effects). 
Luckily no event bursts were reported for the present observations.
SS Aur, with a flux of $\approx 10^{-14}$ergs$~$s$^{-1}$cm$^{-2}$\AA$^{-1}$,
is actually a relatively weak source. 
The minimum acceptable pulse height for ttag FUSE data is controlled by
the parameter PHALOW. Increasing this parameter can reduce the
internal detector background, which is helpful for spectra of very
faint targets (thouh one has to make sure that "real" photon events are not
inadvertently discarded when the treshold is raised). 
Similarly, one can reduce the maximum acceptable pulse height
for ttag FUSE data by reducing the parameter PHAHIGH to a value
closer to the tail of the pulse height distribution (here
also one has to be cautious to make ensure that "real" events
are not removed). 
For SS Aur, we ran CALFUSE after we slightly increased 
PHALOW and decreased PHAHIGH in the parameter files {\bf{scrn*.fit}}
to reduce the noise (background events) as explained above.  

We combined the individual exposures and channels to create a
time-averaged spectrum with a linear, $0.1$ \AA\ dispersion, weightin
the
flux in each output datum by the exposure time and sensitivity of the
input exposure and channel of origin. The details are given here. During,
the observations, Fine Error Sensor A, which images the LiF 1 aperture
was used to guide the telescope. The spectral regions covered by the
spectral channels overlap, and these overlap regions are then used to
renormalize the spectra in the SiC1, LiF2, and SiC2 channels to the flux in
the LiF1 channel. We then produce a final spectrum that covers almost the
full FUSE wavelength range $905-1182$ \AA. The low sensitivity portions of
each channel are discarded.
In most channels there exists a narrow dark stripe of decreased flux
in the spectra running in the dispersion direction. This stripe has been
affectionately known as the "worm" and it can attenuates as much as
50\% of the incident light in the affected portions of the
spectrum. The worm has been observed to move as much as 2000 pixels during a 
single orbit in which the target was stationary, and it appears to be
present in every exposure and, at this time, there is no explanation for it. 
Because of the temporal changes in the strength and position of the worm,
CALFUSE cannot correct target fluxes for its presence. 
Here we take particular care to discard the portion of the spectrum 
where the so-called {\it{worm}} 'crawls', which
deteriorates LiF1 longward of $1125$ \AA\ . Because of this the $1182 
- 1187$ \AA\ region (covered only by the LiF1 channel) is lost. 

We then rescale and combine the spectra. When we
combine, we weight according to the area and exposure time for that
channel and then rebin onto a common wavelength scale with both a $0.5$
\AA\ (Figures 1 \& 2) and a $0.1$ \AA\ (Figures 3 \& 4) resolutions. The 
$0.5$ \AA\ binning is more convenient to identify absorption lines as the
spectra are indeed pretty noisy at $0.1$ \AA\ binning. 

The FUSE spectrum of SS Aur is displayed in figure 1 where we have
identified the strongest absorption features. Table 2 lists in the first
column the central wavelength of the line, second column the flux at that
wavelength, third column the equivalent width (EW) in Angstroms, fourth
column the full-width at half-maximum (FWHM), and the last column the
identified ions.

The FUSE spectrum of RU Peg is displayed in figure 2 and the line
measurements are given in table 3, where the
column headings are the same as for Table 2.
             
Based upon our expectation that the accreting white dwarf is the dominant
source of FUV flux in both systems during quiescence, we carried out a
high gravity photosphere synthetic spectral analysis. The model atmosphere
\citep[TLUSTY]{hub88} , and spectrum synthesis \citep[SYNSPEC]{hub95}
codes and details of our $\chi^{2}_{\nu}$ ($\chi^2$ per degree of freedom)
minimization fitting procedures are discussed in detail in \citet{sio95}
and will not be repeated here.  To estimate physical parameters, we took
the white dwarf photospheric temperature T$_{eff}$, Si and C
abundances, and rotational velocity $v_{rot}$ as free parameters.  We
normalize our fits to 1 solar radius and 1 kiloparsec such that the
distance of a source is computed from $d =
1000(pc)*(R_{wd}/R_{\odot})/\sqrt{S}$, or equivalently the scale factor $S
= \left( \frac{R_{wd}}{R_{\odot}} \right)^2 \left( \frac{d}{kpc}
\right)^{-2}$, is the factor by which the theoretical flux (integrated
over the FUSE wavelength range) has to be multiplied to equal the observed
(integrated) flux.

In preparation for our model fitting of SS Aur, we masked the following
wavelength regions where several narrow emission-like features occur:  
959.5 - 950.0 \AA, 972.4 - 972.6 \AA, 988.6 - 989.0 \AA, 1025.2 - 1026.0
\AA.  For RU Peg, we masked the following wavelength regions : $<$915 \AA,
974 - 980 \AA, 1029 - 1037 \AA, $>$1170 \AA. We chose to vary the
T$_{eff}$, rotational velocity, and silicon and carbon abundances in our
fitting.  The grid of models extended over the following range of
parameters: T$_{eff}/1000$ (K) = 22, 23, ..., 55;
Si = 0.1, 0.2, 0.5, 1.0, 2.0, 5.0; C = 0.1, 0.2, 0.5, 1.0, 2.0, 5.0; and
$v_{rot} \sin{i}$ (km~s$^{-1}$) = 100,200, 400, 600, 800.
Since the distance d = 201 pc from the FGS
parallax, we
used this distance and the reduced $\chi^2$ value to determine the
best-fitting model. In addition, as the WD is expected to be massive
we fixed log $g = 9.0$.

For SS Aur, the best fitting model from our $\chi^2_{\nu}$ 
minimization routine
has the following parameters:  T$_{eff}/1000$ (K) = 33 +15/-7 
Si = 1.0 +1.0/-0.6 times solar, C = 0.1
+0.9/-0.1 times solar, N = 2.0 +1.8/-0.7, $v_{rot}\sin{i} = 400\pm400$
km~s$^{-1}$, $\chi^2_{\nu}$, 
scale factor $= 3.82\times 10^{-4}$. The best-fitting model
is displayed in figure 3. This model gives a reasonable agreement with the
FUSE continuum distribution and lines but yields a distance of 303 pc or
1.5 times the parallax value. Using the above parameters, the N abundance
was found to have an upper limit N$<$8 times solar.

Next, we tried models of accretion disks alone from the grid of Wade and
Hubeny (1998). We fixed the inclination and the white dwarf mass at the 
published values of 41 degrees with $M_{wd} = 1.2 M_{\odot}$. 
The resulting
best-fit had the following parameters: $\chi^2_{\nu}$ = 2.56,
scale factor $=  1.15\times 10^{-2}$ and an accretion rate
$\dot{M} = 10^{-10} M_{\odot}$yr$^{-1}$, corresponding to  a distance
of 931 pc, or 4.6 times the parallax distance. We conclude that 
an optically thick accretion disk by itself does not satisfactorily
account for the FUSE spectrum.

We assessed the effectiveness of combining an accretion disk with a white
dwarf, again fixing the white dwarf mass at $1.2 M_{\odot}$, 
the inclination
angle of the disk at 41 degrees. The white dwarf temperature and disk
accretion rate were free parameters. For this exercise, the best-fitting
combination was the following: $T_{eff}$ = 27,000K, 
$\dot{M} = 1 \times 10^{-11} M_{\odot}$yr$^{-1}$, 
$\chi^2_{\nu}$ = 1.85, scale factor = 5.78 with the accretion disk
accounting for 44\% of the flux and the white dwarf accounting for 56\% of
the flux. However, a model-derived distance is 366 pc, larger than the
parallax distance by a factor of 2.

Finally, we examined whether a two-temperature white dwarf consisting
of a slowly rotating cooler photosphere and a hot rapidly spinning
accretion belt would give better agreement with the FUSE data.
The best-fitting white dwarf plus accretion belt combination had the
following parameters: $T_{eff}$ = 27,000K,  
$T_{belt}$ = 48,000K, $V_{belt}$ = 3000
km/s, $\chi^2_{\nu}$ = 1.74, scale factor = 4.87$ \times 10^{-4}$
with the cooler
photosphere giving 73\% of the FUV light and the accretion belt
providing 27\%. The scale-factor-derived distance is 267 pc, only a factor
of 1.3 larger than the parallax distance.

Thus, based upon the lowest $\chi^2_{\nu}$ 
values achieved and agreement with a
distance of 201 pc, we conclude that either a lone single temperature
white dwarf with Teff = 33,000K or a combined white dwarf with Teff =
27,000K plus an accretion
belt with 
$T_{belt}$ = 48,000K provide the best agreement with the FUSE
spectrum. Models fits with an accretion disk alone or a combined white
dwarf plus an accretion disk imply discrepant distances and larger
$\chi^2$ values and can be eliminated as the source of the FUSE spectrum.

For RU Peg, we also fixed $M_{wd} = 1.2 M_{\odot}$ 
because of its published high mass,  
and fixed log g = 9.0. The FGS
parallax of Johnson et al. gives a distance d = 282 pc. We used this
distance and the reduced $\chi^2_{\nu}$ 
value to distinguish which model is the
best-fitting among a single temperature white dwarf alone, a steady
state, optically thick accretion disk alone, a combined photosphere plus
an accretion disk or a two-temperature white dwarf consisting of a cooler,
more slowly rotating photosphere and a hot, rapidly spinning accretion
belt.

For RU Peg, the best fitting model from our $\chi^2_{\nu}$ minimization
routine (with 3 sigma error bars) has the following parameters:
T$_{eff}/1000$ (K) = 53 +6/-7, Si = 0.1 +1.0/-0.1
times solar, C = 0.1 +0.9/-0.1 times solar, an upper limit N abundance of
N $<$ 8.0 times solar, $v_{rot}\sin{i}$ = 100 +400/-100 km~s$^{-1}$,
$\chi^2_{\nu}$ = 4.06 , and scale factor $= 5.53\times 10^{-4}$.
This best-fit gave a distance of 263 pc, within about 20 parsecs of the
parallax distance. The best-fitting model is displayed in figure 4.
Unfortunately, the RU Peg spectrum is under-exposed and therefore the S/N
and spectrum quality is less than satisfactory.

For an accretion disk alone, we fixed the inclination angle at 41 degrees,
$M_{wd} = 1.2 M_{\odot}$, 
with the accretion rate a free parameter. We found the
best-fitting disk model to have $\dot{M}=10^{-9} M_{\odot}$yr$^{-1}$ 
with a $\chi^2_{\nu}$
= 4.18, a scale factor $= 5.57\times 10^{-3}$ but a distance of 1.34
kpc, clearly erroneous. When we combined white dwarf plus accretion disk
models with $T_{eff}$ varying between 36,000 and 60,000K (in steps of 1000K)  
and $\dot{M}$ varying over the full range of the disk model grid, there was no
improvement in the fitting and the scale factor-derived distance was still
grossly too large.

Next, we tested two-temperature fits with a cooler, slowly rotating
photosphere and a hot, rapidly spinning, equatorial belt as expected from
disk accretion. In this experiment, we varied the WD $T_{eff}$ 
between 42,000K and 60,00K, the accretion belt temperature between 50,000K
and 60,000K in steps of 1000K and kept the C and Si abundances of the
white dwarf fixed. Once again, as in the white dwarf plus disk case, there
was no improvement in the $\chi^2_{\nu}$ value.
 
\section{Results and Discussions}

The quality of the model fits to the FUSE spectra of the two systems is
quite different. The fit to SS Aur is very much in agreement with a model
white dwarf atmosphere with log $g$ = 9.0 and $T_{eff}$ = 33,000K.
This fit to the FUSE spectrum provides independent confirmation of the
results of \citet{lak01} who also found that the far UV IUE spectra were
dominated by a hot, massive white dwarf. The $T_{eff}$ they derived
with IUE
for the white dwarf in SS Aur was 30,000K. This is surprising because it
was widely felt that the white dwarf in SS Aur was not exposed, the system
was disk-dominated in the far UV and could not be analyzed unambiguously.

In SS Aur, it is also highly significant that there is little evidence of
an additional hot component other than a single temperature white dwarf
photosphere. The absence of C\,{\sc iii} (1175 \AA) absorption
in the FUSE
spectrum suggests the possibility that the white dwarf is deficient in
carbon. If so, this could be an indication that past thermonuclear
processing (ancient novae) depleted the carbon. This possibility is
supported by the indication that the N-abundance in the SS Aur WD surface
layers is elevated above solar. An alternative picture discussed by 
\citet{gan03} 
suggest that the N/C anomaly seen in the dwarf
novae BZ UMa, EY Cyg, 1RXS J232953.9+062814, and now CH UMa 
\citep{dul02,dul04} 
may have its origin in a CV with an originally more massive
donor star ($M_{2} > 1.5 M_{\odot}$) which survived thermal time scale mass
transfer (\citet{sch02} and references therein). In such a system,
the white dwarf would be accreting from the peeled away CNO-processed core
stripped of its outer layers during the thermal timescale mass transfer.

Our FUSE spectrum of RU Peg likewise reveals a very hot white dwarf in
agreement with the analysis of the IUE archival spectra of RU Peg in
quiescence. We find that $T_{eff}$ = 49,000K for the white dwarf, is very
close to the $T_{eff}$ derived by \citet{sio02}. However, the FUSE
spectrum appears more complex than for SS Aur. There is evidence for a hot
component. The best-fitting single temperature, high gravity, solar
composition white dwarf models reveal a 49,000K white dwarf as the
dominant source of the FUV continuum. The scale factors for the hot white
dwarf fits yield distances of 230 pc and 260 pc, the latter value lying
within the range of uncertainty of the new FGS parallax. We note however
that although a hot single-temperature (50,000K) white dwarf agrees best
with the far UV observations, we cannot rule out that the far UV continuum
could be produced by a cooler, slowly rotating white dwarf and a rapidly
spinning, very hot accretion belt covering a small fraction of the white
dwarf surface but providing the vast majority of the FUV flux.

\section{Acknowledgements}
This research was carried out with the support of
NASA through FUSE grant NAG5-12067, 
and by NSF grant AST99-01955, both to Villanova University.

{}

\clearpage 
     
\begin{deluxetable}{lcc}

\tablewidth{0pt}
\tablecaption{System Parameters \label{tab:ssaur}}
\tablehead{\colhead{Object}&\colhead{RU PEG}&\colhead{SS AUR}}
\startdata

Orbital Period (days) & 0.3746 & 0.1828\\
V Magnitude (quiescence) & 12.6 & 14.5\\
V Magnitude (outburst) & 9.0 & 10.5\\
Recurrence Time (days) & 75-85 & 40-75\\
Distance (pc) & 282 & 201 \\ 

\enddata

\end{deluxetable}

\newpage
     
\begin{deluxetable}{cccccc}

\tablewidth{0pt}
\tablecaption{SS Aur FUSE Line Measurements}
\tablehead{Center & Flux                             & EW & FWHM & Line & Identification \\ 
 $<$ \AA\ $>$ & $<$ erg/s/cm$^2$/\AA\ $>$ & $<$ \AA\ $>$ & $<$ \AA\ $>$ & & $<$ \AA\ $>$  
}
\startdata
 910.9  &  -5.6e-15    &   0.624 & 2.055 & S\,{\sc iii} & 911.7\\ 
 916.7  &  -4.9e-15    &   0.680 & 0.628 & Fe\,{\sc ii} & 915.97\\ 
        &              &         &       & N\,{\sc ii}  & 915.6,
916.0, 916.7 \\ 
 921.5  &  -1.7e-14    &   2.297 & 3.166 & N\,{\sc iv}  & 922.0,
922.5, 923.2 \\ 
 927.5  &  -3.4e-15    &   0.644 & 1.301 & ?     &                      \\ 
 929.9  &  -3.8e-15    &   0.649 & 0.734 & He\,{\sc ii} & 930.3\\ 
 952.6  &  -1.9e-14    &   3.303 & 4.169 & ? P\,{\sc iv}& 950.7 ?\\ 
 964.5  &  -1.7e-14    &   2.259 & 3.315 & ?     &                      \\ 
 968.5  &  -1.9e-14    &   2.506 & 4.717 & ?     &                      \\ 
 980.6  &  -3.6e-14    &   2.589 & 3.844 & N\,{\sc iii} & 979.9, 980.0\\ 
 985.9  &  -6.5e-14    &   4.771 & 3.015 & Cl\,{\sc iv} & 985.0,
985.8, 986.1 \\ 
 994.0  &  -5.3e-14    &   2.862 & 6.698 & Fe\,{\sc iii}& 993.1,
994.7, 995.1 \\ 
 1003.1 &  -1.9e-14    &   1.260 & 3.268 & N\,{\sc iii} & 1002.85,
1003.21, \\ 
        &              &         &       & O\,{\sc iii} & 1003.35\\ 
 1010.4 &  -9.1e-15    &   0.788 & 1.995 & C\,{\sc ii}  & 1010.1,
1010.4, \\ 
        &              &         &       & O\,{\sc iii}& 1010.5\\ 
 1014.6 &  -2.9e-14    &   2.522 & 4.092 & Cl\,{\sc iii}& 1015.0,\\ 
        &              &         &       & S\,{\sc iii}& 1015.5,
1015.8 \\ 
 1024.5 &  -1.2e-13    &   10.98 & 17.35 & He\,{\sc ii} & 1025.4\\ 
 1029.3 &  -1.5e-14    &   1.281 & 2.755 & P\,{\sc iv}  & 1030.5\\ 
 1037.2 &  -1.8e-14    &   1.272 & 2.141 & C\,{\sc ii}  & 1036.34,
1037.02 \\ 
 1065.7 &  -6.3e-14    &   2.839 & 5.751 & S\,{\sc iv}  & 1066.6,\\ 
        &              &         &       & O\,{\sc iv}  & 1067.8,\\ 
        &              &         &       & Si\,{\sc iv} & 1066.6\\ 
 1078.0 &  -2.7e-14    &   1.196 & 3.261 & S\,{\sc iii} & 1077.14,\\ 
        &              &         &       & N\,{\sc iv}  & 1078.71\\ 
 1085.7 &  -2.7e-14    &   1.213 & 5.331 & N\,{\sc ii}  & 1085.5,
1085.7 \\ 
 1092.8 &  -2.2e-14    &   0.977 & 2.438 & ?     &                      \\ 
 1109.3 &  -2.8e-14    &   1.393 & 2.889 & Si\,{\sc iii}& 1109.94,
1109.97 \\ 
 1113.3 &  -1.8e-14    &   0.946 & 3.412 & Si\,{\sc iii}& 1113.20, 
1123.23 \\ 
 1122.7 &  -1.6e-14    &   0.820 & 3.470 & Si\,{\sc iv} & 1122.50\\ 
 1128.8 &  -1.7e-14    &   0.887 & 2.770 & Si\,{\sc iv} & 1128.34\\ 
 1144.0 &  -4.9e-14    &   2.434 & 5.765 & Si\,{\sc iii}& 1144.3\\ 

\enddata

\end{deluxetable}

\newpage
     
\begin{deluxetable}{cccccc}

\tablewidth{0pt}
\tablecaption{RU Peg FUSE Line Measurements}
\tablehead{Center & Flux & EW & FWHM & Line & Identification  \\ 
 $<$ \AA\ $>$ & $<$ erg/s/cm$^2$/\AA\ $>$ & $<$ \AA\ $>$ & $<$ \AA\ $>$ & & $<$ \AA\ $>$  
}
\startdata
 912.40 & -4.2e-13   &    6.941 & 9.226 & S\,{\sc iii}   & 911.74?                   \\ 
 923.44 & -3.7e-13   &    1.19  & 2.361 & N\,{\sc iv}    & 923.22,923.68             \\ 
 930.58 & -1.7e-13   &    1.46  & 2.197 & He\,{\sc ii}   & 930.34                    \\ 
 933.84 & -9.7e-14   &    0.856 & 1.067 & He\,{\sc ii}   & 933.40                    \\ 
 938.63 & -1.4e-13   &    1.202 & 2.208 & He\,{\sc ii}   & 937.39                    \\ 
 965.25 & -1.3e-13   &    3.90  & 5.782 & Fe\,{\sc iii}  & 967.19 ?                  \\ 
 972.49 & -1.4e-14   &    1.706 & 3.06  & He\,{\sc ii}   & 972.11                    \\ 
 986.12 & -1.7e-13   &    1.436 & 2.221 & Cl\,{\sc iv}   & 984.95,985.75,986.09      \\ 
 992.15 & -1.2e-13   &    1.116 & 1.646 & N\,{\sc iii}   & 991.58,                   \\ 
        &            &          &       & He\,{\sc ii}   & 992.36                    \\ 
 1003.3 & -1.6e-13   &    1.477 & 3.109 & N\,{\sc iii}   & 1002.85, 1003.21,         \\ 
        &            &          &       & O\,{\sc iii}   & 1003.35                   \\ 
 1008.9 & -5.2e-14   &    0.523 & 0.646 & O\,{\sc iii}   & 1008.10,1008.39, 1008.97  \\ 
        &            &          &       & Cl\,{\sc iii}  & 1008.78                   \\ 
 1013.6 & -9.3e-14   &    0.981 & 2.312 & ?       &                           \\
 1025.4 & -2.1e-13   &    1.622 & 1.692 & He\,{\sc ii}   & 1025.36                   \\ 
 1031.9 & -2.4e-13   &    1.197 & 2.027 & P\,{\sc iv}    & 1030.5, 1033.11           \\ 
 1037.3 & -2.8e-13   &    1.7   & 2.082 & C\,{\sc ii}    & 1036.34, 1037.02          \\ 
 1050.4 & -8.2e-13   &    6.349 & 11.98 & O\,{\sc iii}   & 1050.4                    \\ 
 1063.0 & -1.3e-13   &    1.117 & 1.418 & S\,{\sc iv}    & 1062.68                   \\ 
 1073.5 & -9.0e-14   &    0.706 & 1.405 & S\,{\sc iv}    & 1072.99, 1073.52          \\ 
 1078.2 & -2.1e-13   &    1.676 & 3.342 & S\,{\sc iii}   & 1077.14,                  \\ 
        &            &          &       & N\,{\sc iv}    & 1078.71                   \\ 
 1096.4 & -3.5e-13   &    5.321 & 14.0  & ?       &                           \\ 
 1109.3 & -1.1e-13   &    1.076 & 2.94  & Si\,{\sc iii}  & 1109.94, 1109.97          \\ 
 1113.6 & -7.0e-14   &    0.686 & 1.883 & Si\,{\sc iii}  & 1113.20, 1123.23          \\ 
 1160.6 & -8.9e-14   &    11.17 & 23.19 & O\,{\sc iii}   & 1160.18                   \\ 
 1176.1 & -1.6e-13   &    1.19  & 2.361 & C\,{\sc iii}   & 1174.93, 1175.26,         \\ 
        &            &          &       &         & 1175.71,1175.99, 1176.37  \\ 

\enddata

\end{deluxetable}

\clearpage

\begin{figure}
\plotone{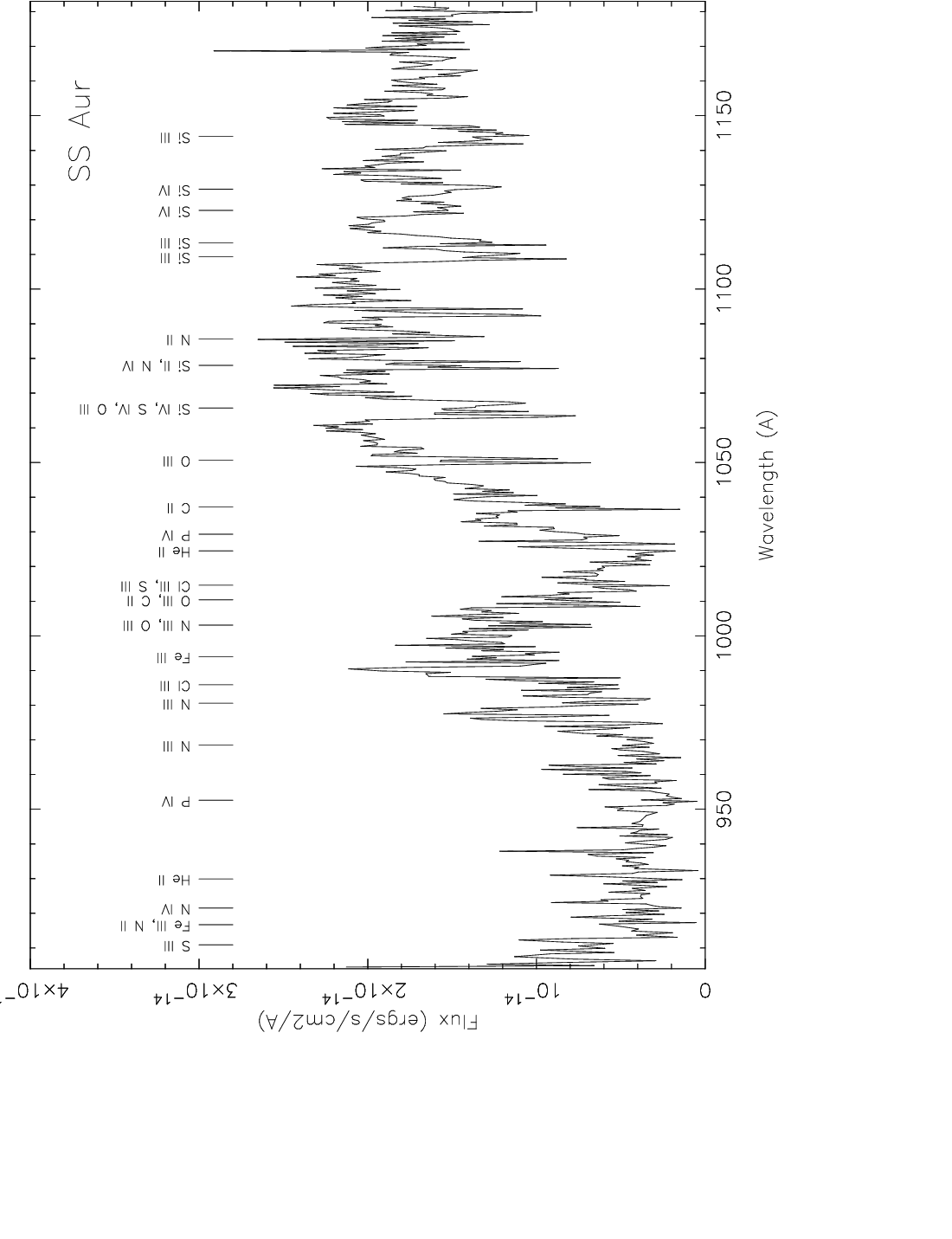}
\figurenum{1}
\vspace{-3.cm} 
\caption{Line identification of the FUSE spectrum of SS Aur
observed on February 13, 2002 (MJD52318). 
in quiescence, 28 days after outburst.  
The time-averaged spectrum was generated by combining 
the individual exposures and channels. The total exposure time was 
14,513 s, through the FUSE LWRS aperture.  
The spectrum is binned here at 0.5 \AA\ for clarity. }
\end{figure}

\clearpage

\begin{figure}
\plotone{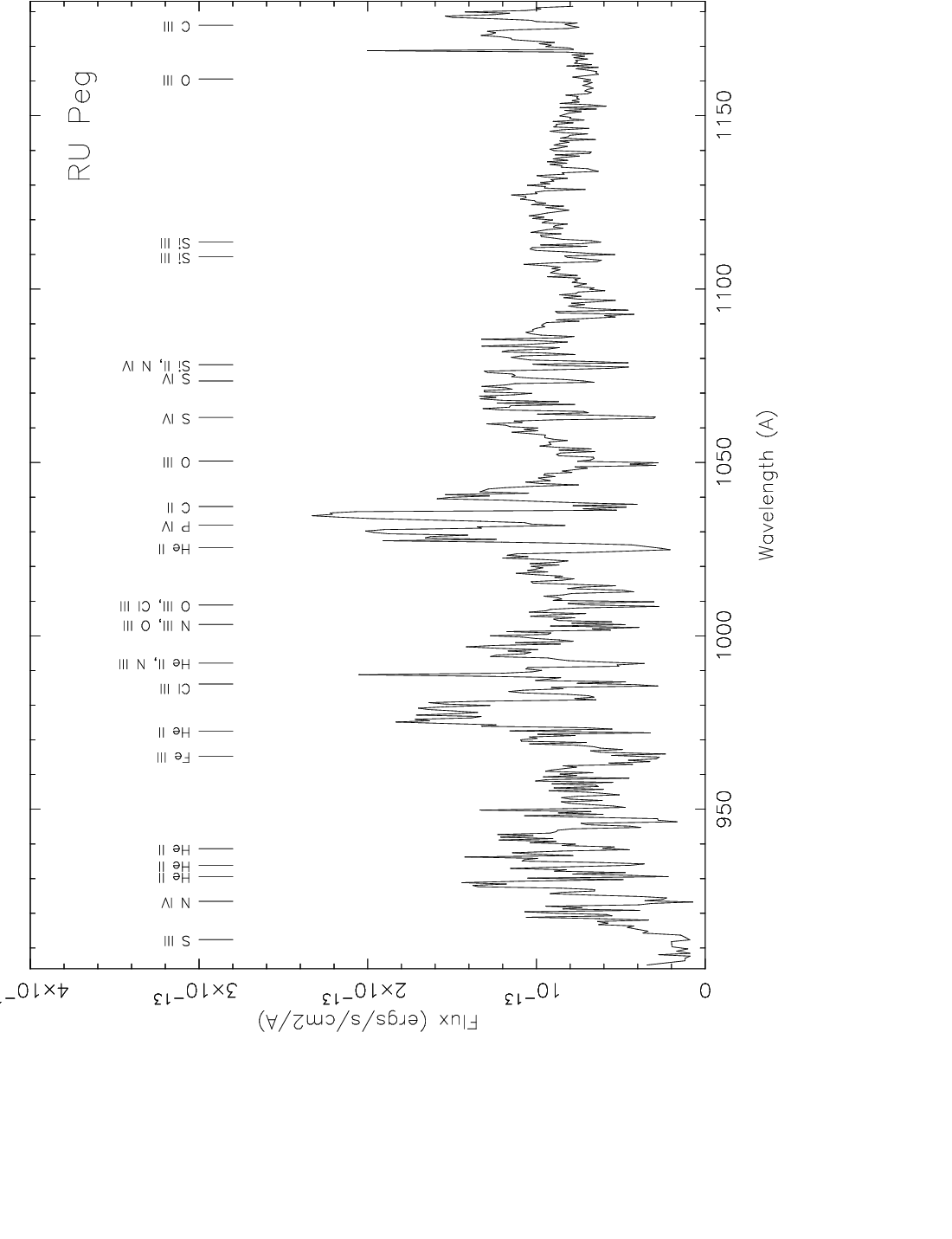}
\figurenum{2}
\vspace{-3.cm} 
\caption{Line identification of the FUSE spectrum of RU Peg
observed on July 4th, 2002 (MJD52459). 
in quiescence, 60 days after outburst.  
The time-averaged spectrum was generated from a  
single exposure by combining the different channels. 
The total exposure time was 1,060 s, through the FUSE LWRS aperture.  
The spectrum is binned here at 0.5 \AA\ for clarity. }
\end{figure}

\clearpage

\begin{figure}
\plotone{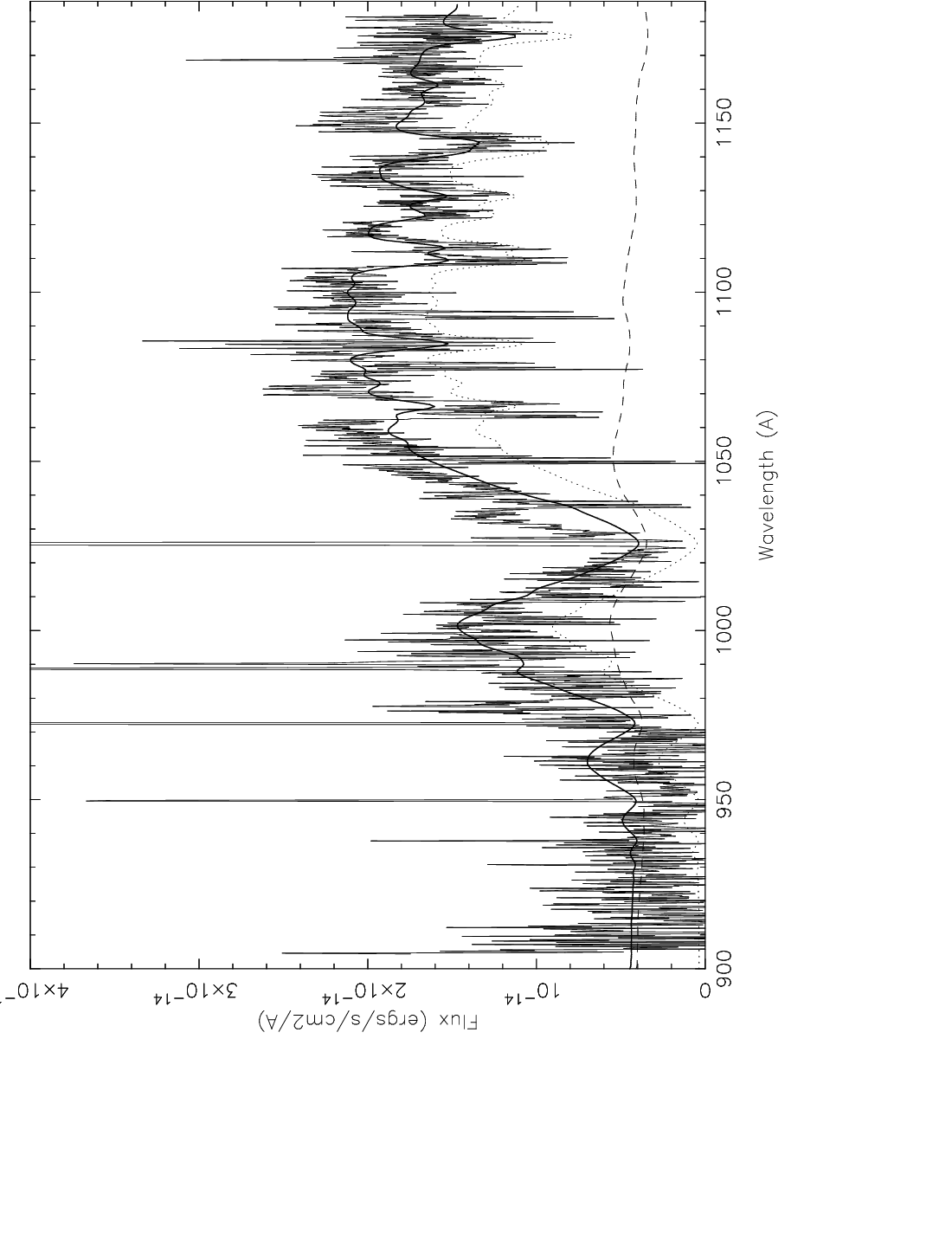}
\figurenum{3}
\vspace{-3.cm} 
\caption{FUSE spectrum of SS Aur with the best fit model. Here the 
spectrum is binned at 0.1 \AA\. The best fit model consists of a white
dwarf atmosphere with $T_{eff}=31,000$K.}
\end{figure}

\clearpage

\begin{figure}
\plotone{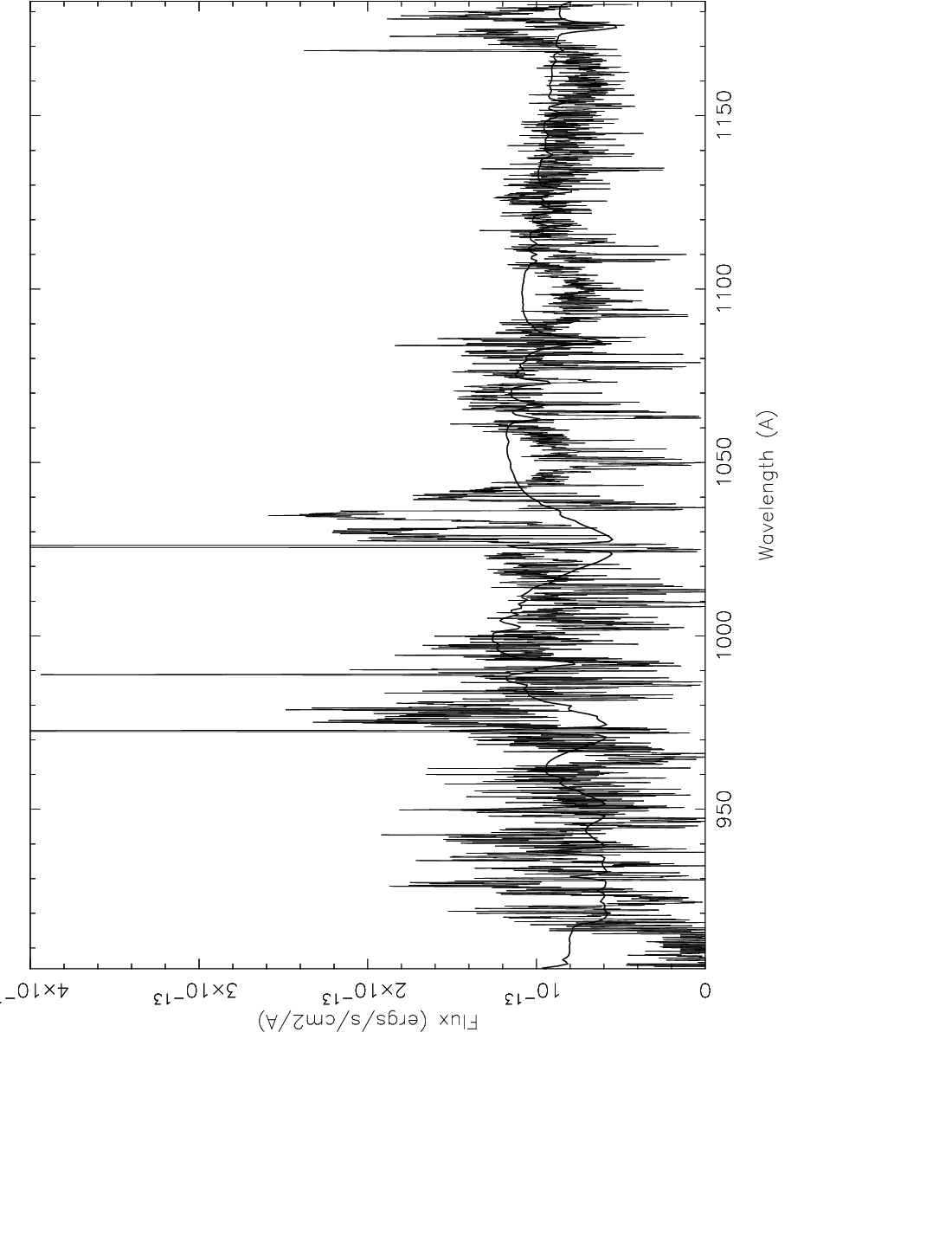}
\figurenum{4}
\vspace{-3.cm} 
\caption{FUSE spectrum of RU Peg with the best fit model. Here the 
spectrum is binned at 0.1 \AA\. The best fit model consists of a white
dwarf atmosphere with $T_{eff}=49,000$K.}
\end{figure}

\end{document}